\PassOptionsToPackage{pdfa}{hyperref}
\documentclass[9pt,
    sigconf,
    anonymous=false, 
    timestamp=false, 
    balance=true, 
    nonacm=true, 
    screen=true, 
    authordraft=false, 
    review=false]{acmart}		%

\usepackage[a-2b]{pdfx}

\input{datalabmacros}

\setcopyright{none}

\newcommand\vldbdoi{10.14778/3827998.3828153}
\newcommand\vldbpages{4932-4938}
\newcommand\vldbvolume{19}
\newcommand\vldbissue{12}
\newcommand\vldbyear{2026}
\newcommand\vldbauthors{\authors}
\newcommand\vldbtitle{\shorttitle} 
\newcommand\vldbavailabilityurl{https://northeastern-datalab.github.io/relational-language-tutorial/}
\newcommand\vldbpagestyle{empty}

\begin{document}

\newcommand{\ourTitle}{An Extended Tutorial and Vocabulary for Relational Language Design in an Era of AI-Assisted Query Generation}
\title[\ourTitle]{\ourTitle}

\author{Wolfgang Gatterbauer}
\orcid{0000-0002-9614-0504}
\affiliation{%
    \orcidicon{0000-0002-9614-0504}
    \institution{Northeastern University}
    \city{Boston}
    \state{MA}
    \country{USA}
}

\begin{abstract}
Relational query languages have been studied and used for more than 50 years, with SQL dominant in practice.
Today, queries are increasingly generated by machines and read by humans.
At the same time, the landscape also includes dataframe, pipeline, logical, functional, graph, and relational programming notations.
These developments invite two related questions beyond expressive power: 
\emph{which relational structures do languages make explicit}, and 
how well can notation \emph{support users in reading and revising queries}?

This 3-hour tutorial extends an earlier SIGMOD'26 tutorial in three directions: 
recursive and path queries (connecting relational and graph query languages), 
nested relational data, 
and relational languages for problems beyond PTIME. 
Rather than beginning from formal definitions, 
we start from example queries and compare how different languages express the same intent. 
To compare recurring structure across notations, 
we use Abstract Relational Calculus (ARC) and Relational Diagrams as reference representations.

From these examples, we develop a vocabulary for relational language design, 
including \emph{information need}, 
\emph{query mapping}, \emph{relational pattern structure}, \emph{relational pattern denotation}, 
and \emph{semantic conventions}. 
Participants will leave with a framework for comparing existing and future relational languages, 
a precise vocabulary for articulating design trade-offs, 
and a concrete set of examples connecting classical database languages with alternative proposals.

\end{abstract}

\maketitle

\pagestyle{\vldbpagestyle}
\begingroup\small\noindent\raggedright\textbf{PVLDB Reference Format:}\\
\vldbauthors. \vldbtitle. PVLDB, \vldbvolume(\vldbissue): \vldbpages, \vldbyear.\\
\href{https://doi.org/\vldbdoi}{doi:\vldbdoi}
\endgroup
\begingroup
\renewcommand\thefootnote{}\footnote{\noindent
This work is licensed under the Creative Commons BY-NC-ND 4.0 International License. Visit \url{https://creativecommons.org/licenses/by-nc-nd/4.0/} to view a copy of this license. For any use beyond those covered by this license, obtain permission by emailing \href{mailto:info@vldb.org}{info@vldb.org}. Copyright is held by the owner/author(s). Publication rights licensed to the VLDB Endowment. \\
\raggedright Proceedings of the VLDB Endowment, Vol. \vldbvolume, No. \vldbissue\ %
ISSN 2150-8097. \\
\href{https://doi.org/\vldbdoi}{doi:\vldbdoi} \\
}\addtocounter{footnote}{-1}\endgroup

\ifdefempty{\vldbavailabilityurl}{}{
\vspace{.3cm}
\begingroup\small\noindent\raggedright\textbf{PVLDB Artifact Availability:}\\
The slides from the tutorial will be made available on the tutorial web page: \url{\vldbavailabilityurl}.
\endgroup
}

\section{Introduction}

Donald Knuth is quoted as saying,
``\emph{I think of a programming language as a tool to convert a programmer's
mental images into precise operations that a machine can perform}''
~\cite{Knuth:interview:2013}.
In the same spirit, the database community has primarily viewed relational
query languages as specifications for translating information needs
into executable commands.
This perspective has helped make \emph{expressive power} a central criterion
for comparing and designing relational languages~\cite{DBLP:persons/Codd72}.
Yet \emph{what a language can say} is only one aspect of language design.
Another one is \emph{how effectively} it lets us formulate, understand, and revise what we want to say.

In his 1979 Turing Award lecture
``\emph{Notation as a Tool of Thought}''~\cite{Iverson:1980},
Iverson cites Whitehead via Cajori:
``\emph{By relieving the brain of all unnecessary work, a good notation sets
it free to concentrate on more advanced problems, and in effect increases
(our) mental power...}''
Good notation can make relational structure explicit
(e.g., a join path) 
so that readers can perceive it directly rather than having to reconstruct and
retain it in their minds as they read a query.
Yet this view of notation design for ``cognition enhancement'' has not received much attention 
in relational language design.
It echoes the idea of \emph{linguistic relativity}: the
languages we use can shape how we represent and reason about
problems~\cite{wiki:linguistic_relativity}.

This perspective is especially timely as generative AI and large language
models (LLMs) are changing how users interact with data.
Queries are no longer only instructions for machines.
Increasingly, they are
shared working representations through which \emph{humans can understand what
machines have done or intend to do}:
machines generate and transform queries, while humans inspect, interpret,
verify, and revise them~\cite[Fig.~2]{Gatterbauer2022PrinciplesQueryVisualization}.
As noted in
\cite{DBLP:journals/corr/abs-2504-11259,
Gatterbauer2022PrinciplesQueryVisualization,
CIDR:2026:ARC},
this shifts attention from human query composition toward machine query
generation and \emph{human query interpretation}.
When machine-generated queries can be wrong or misleading,
making them easy to read, verify, and revise becomes crucial.

At the same time, there is renewed interest in relational language design.
A number of recent efforts propose extensions or alternatives to
SQL~(including
\cite{DBLP:conf/cidr/0001L24,
DBLP:journals/pvldb/ShuteBBBDKLMMSWWY24,
DBLP:conf/sigmod/ArefGKLMMMMNPRS25,
prql})
and differ 
in how they
present relational structure: through nested or pipelined composition,
named or positional access to relation components, and tuple or domain
variables.
Such choices affect which structural patterns are easy to recognize and
which revisions are easy to make.
Yet comparing these choices systematically requires a common vocabulary
that \emph{separates notation from relational structure and semantics}.

\textbf{What we need}, more broadly, is systematic research about how design choices in
relational languages affect how
humans and machines think and reason. 
A necessary ingredient is a framework and vocabulary
for discussing \emph{how different relational languages realize the same relational intent (or query mapping) while making different structures, assumptions, and notations visible} to humans and machines.
Without a precise vocabulary, debates about relational languages tend to
confuse surface syntax and semantic conventions.

\begin{figure}[t]
    \centering        
    \includegraphics[scale=0.38]{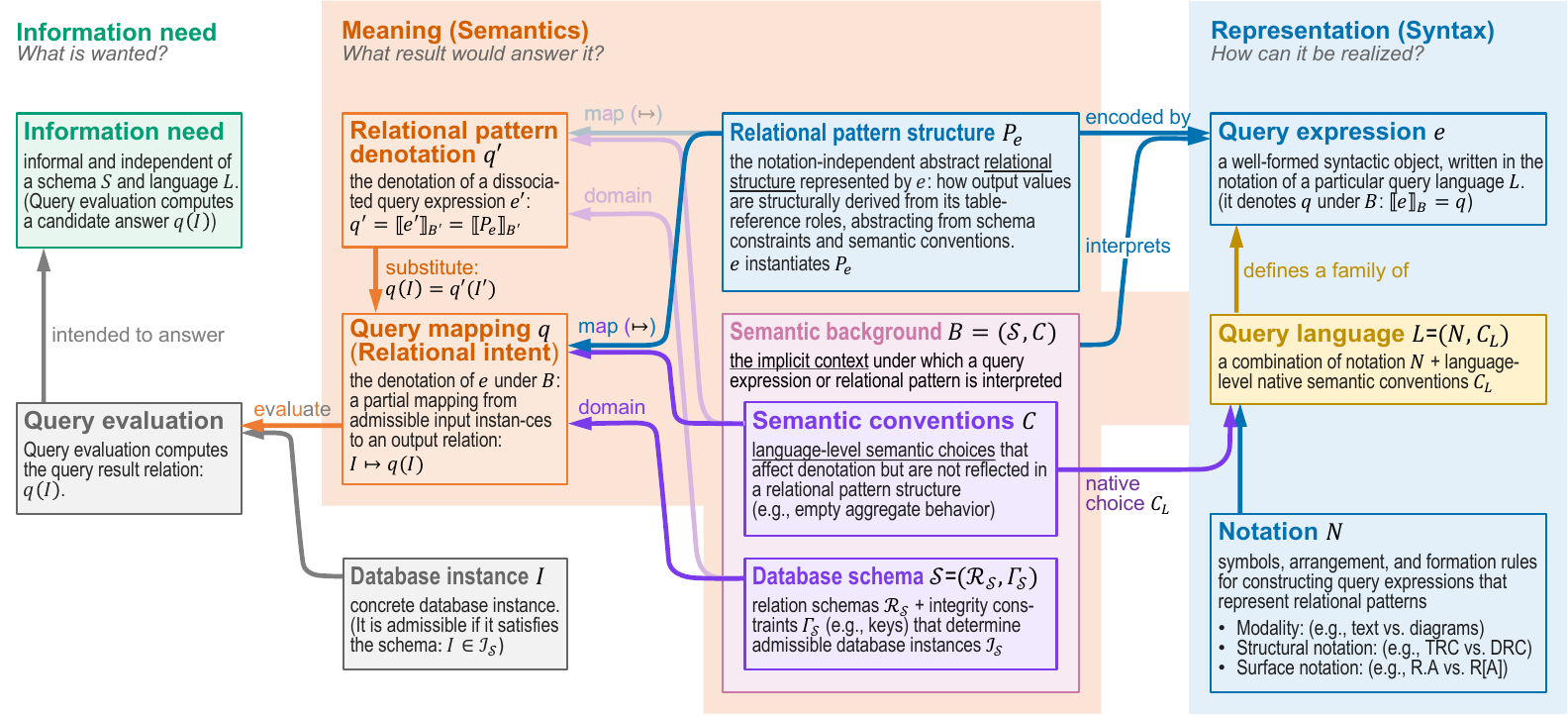}
    \caption{Relationships between concepts in the vocabulary of our relational language design framework 
    that we will use.}
    \label{Fig_Vocabulary}
\end{figure}

\textbf{This tutorial}
develops and applies such a relational language design
framework~(\cref{Fig_Vocabulary})
that distinguishes and disentangles
several concepts
that discussions of query languages often conflate. 
Rather than beginning with formal definitions, we start from a fixed set of
queries and compare their realizations across various relational languages.
The examples expose differences in relational pattern, semantic conventions,
and structural notation before these distinctions are formalized.
To make recurring relational patterns explicit, we use Abstract Relational
Calculus (ARC)~\cite{CIDR:2026:ARC}, together with
\diagrams~\cite{DBLP:journals/sigmod/GatterbauerD25,10.1145/3639316},
as common reference representations.
Our goal is to survey languages, clarify the design space,
make the comparison concrete through a set of recurring examples,
and develop a precise vocabulary for relational language design
to enable clear communication.

This 3-hour tutorial extends an earlier SIGMOD 2026
tutorial~\cite{gatterbauer2026tutorial} in three directions:
(1) recursion and path queries (connecting relational and graph query
languages),
(2) nested relational data, and
(3) relational formalisms for problems beyond PTIME.

\section{An illustrative end-to-end example}
\label{sec:intro:example}

\begin{quote}
\small\itshape
``A good stock of examples, as large as possible, is indispensable
for a thorough understanding of any concept, and when I want to
learn something new, I make it my first job to build one.''
\par
\hfill\normalfont
\textemdash\ Paul Halmos~\cite[p.~63]{halmos:1985}
\end{quote}

\introparagraph{Information need, query expressions, and query mappings}
An \emph{\HA{information need}} $\eta$ describes the information sought
in a given context, often informally, such as
``\emph{Return the sum of each employee's sales}.''
It does not yet fix a database schema or query language, nor necessarily
all semantic details, such as how employees without sales should be
treated.

\Cref{fig:intro:instance}
shows a \HA{database instance} $I_1$ over a
\HL{database schema} $\mathcal{S}_1$ with
\HA{relation names}[relation name]
$\mathsf{Empl}$ and $\mathsf{Sales}$, whose current
\HA{relation values}[relation value] $I_1(\mathsf{Empl})$ and
$I_1(\mathsf{Sales})$ each contain three tuples.
The schema also specifies \HA{integrity constraints}, which restrict
the admissible database instances:
the primary keys are
$\{\mathsf{eid}\}$ for $\mathsf{Empl}$ and
$\{\mathsf{eid},\mathsf{date}\}$ for $\mathsf{Sales}$, and
$\mathsf{Sales}.\mathsf{val}$ is \texttt{NOT NULL}.

\Cref{fig:intro:query:sql:e1,fig:intro:query:souffle:e2}
show two
\emph{\HA{query expressions}[query expression]} for answering
$\eta$: an \SQL expression $e_1$ and a \souffle expression $e_2$.
To interpret a query expression $e$, we fix a
\emph{\HA{semantic background}}
$B=(\mathcal{S},C)$
where the \HL{database schema} $\mathcal{S}$ determines the admissible input
instances and the \emph{\HA{semantic conventions}} $C$ determine how the constructs
in $e$ are interpreted. The resulting
\emph{\HA{query mapping}} is the {denotation}
\[
  q_{e,B}
  \coloneqq
  \llbracket e\rrbracket_B,
\]
which we write as $q$ if $e$ and $B$ are clear from context.
It is a possibly partial \emph{function} from admissible instances of
$\mathcal{S}$ to output \HL{relation values}[relation value], and
$q_{e,B}(I)$ is the result of evaluating $e$ on $I$ under $B$.%
\footnote{The earlier SIGMOD'26 tutorial~\cite{gatterbauer2026tutorial}
used ``query intent'' for what we now call \emph{information need} and
``relational intent'' for \emph{query mapping}. The original terminology
was meant to contrast an informal information need with its formal
realization as a mapping, but feedback showed that the two
uses of ``intent'' were confusing. ``Query mapping'' now also follows the
terminology of Abiteboul et al.~\cite[Ch.~4]{DBLP:books/aw/AbiteboulHV95}
and is used in addition to ``relational intent.''
``Information need'' is preferred over ``user intent'' because it does not
assume a human user.
}

A \HA{query language} $L=(N,C_L)$ consists of a
\HL{notation} $N$ together with its native semantic conventions $C_L$.%
\footnote{%
\emph{\HA{Notation}[notation]} 
is the externally represented scheme of symbols, arrangement conventions, and formation rules by which query expressions are constructed and recognized.
We distinguish three components:
\emph{\HA{modality}[modality]} (the external representational form in which notation is realized, such as text or diagrams),
and two types of notational design choices,
\emph{structural notation}
and
\emph{surface notation},
which are either modality-independent or modality-dependent.
}
An expression written in $L$ is therefore interpreted natively using
$C=C_L$, although we may also interpret it under another compatible
choice of $C$.

\begin{figure}[t]
\setlength{\tabcolsep}{3pt}
\renewcommand{\arraystretch}{0.92}
\centering

\begin{subfigure}[t]{0.66\linewidth}
  \centering
  \begin{tabular}[t]{|c|c|}
    \multicolumn{2}{l}{$\textsf{Empl}$} \\
    \hline
    \rowcolor{gray!15}
    \underline{\textsf{eid}} & \textsf{k} \\
    \hline
    \texttt{a} & 1 \\
    \texttt{b} & 2 \\
    \texttt{c} & 3 \\
    \hline
  \end{tabular}
  \hspace{1em}
  \begin{tabular}[t]{|c|c|c|}
    \multicolumn{3}{l}{$\textsf{Sales}$} \\
    \hline
    \rowcolor{gray!15}
    \underline{\textsf{eid}} & \textsf{val} & \underline{\textsf{date}} \\
    \hline
    \texttt{a} & 10 & \texttt{1/1/26} \\
    \texttt{a} & 20 & \texttt{1/2/26} \\
    \texttt{b} & 40 & \texttt{1/1/26} \\
    \hline
  \end{tabular}
  \caption{$\mathcal{S}_1$, $I_1$}
  \label{fig:intro:instance}
\end{subfigure}

\begin{subfigure}[t]{\linewidth}
  \centering
  \vspace*{4pt}
  \begin{minipage}[t]{39mm}
\begin{lstlisting}[style=sqlcode]
select E.eid,
 (select sum(S.val) as sm
  from Sales S
  where E.eid = S.eid)
from Empl E
\end{lstlisting}
\vspace{-2mm}
  \end{minipage}
  \caption{$e_1$ (SQL)}
  \label{fig:intro:query:sql:e1}
\end{subfigure}

\begin{subfigure}[t]{\linewidth}
  \centering
  \begin{minipage}[t][4mm][t]{0.96\linewidth}
    \vspace*{5pt}
    \centering
    \begin{tabular}{@{}l@{}}
      \texttt{Q(e, sum v: \{Sales(e,v,\_)\}) :- Empl(e,\_).}
    \end{tabular}
  \end{minipage}
  \caption{$e_2$ (Souffl\'e)}
  \label{fig:intro:query:souffle:e2}
  \vspace{2mm}
\end{subfigure}

\begin{subfigure}[t]{0.3\linewidth}
  \centering
  \begin{minipage}[t][14mm][t]{\linewidth}
    \centering
    \begin{tabular}[t]{|c|c|}
      \hline
      \rowcolor{gray!15}
      \textsf{eid} & \textsf{sm} \\
      \hline
      \texttt{a} & 30 \\
      \texttt{b} & 40 \\
      \texttt{c} & \texttt{NULL} \\
      \hline
    \end{tabular}
  \end{minipage}
  \caption{$q_{e_1,(\mathcal{S}_1,C_{\SQL})}(I_1)$
  }
  \label{fig:intro:out:sql:e1}
\end{subfigure}%
\hspace{10mm}
\begin{subfigure}[t]{0.3\linewidth}
  \centering
  \begin{minipage}[t][14mm][t]{\linewidth}
    \centering
    \begin{tabular}[t]{|c|c|}
      \hline
      \rowcolor{gray!15}
      \textsf{eid} & \textsf{sm} \\
      \hline
      \texttt{a} & 30 \\
      \texttt{b} & 40 \\
      \texttt{c} & 0 \\
      \hline
    \end{tabular}
  \end{minipage}
  \caption{$q_{e_2,(\mathcal{S}_1,C_{\souffle})}(I_1)$}
  \label{fig:intro:out:souffle:e2}
\end{subfigure}

\caption{A database schema and instance with underlined attributes forming
primary keys (a), 
two expressions with similar conceptual evaluation strategies
intended to answer the same information need
(b, c), and their different query results
under the languages' native semantic
conventions (d, e).}
\label{fig:intro:motivating}
\end{figure}

\introparagraph{Semantic conventions affect query mappings}
Despite their different surface notations, the \emph{\HA{conceptual evaluation strategies}[conceptual evaluation strategy]} of $e_1$ and $e_2$ share the same relational core:
each takes an employee tuple from $\mathsf{Empl}$,
selects the matching $\mathsf{Sales}$ tuples on $\mathsf{eid}$, sums
their $\mathsf{val}$ values, and returns the employee identifier
together with the sum.%
\footnote{For a description of \SQL's conceptual evaluation strategy, see \cite[Sect.~5.2 and 5.4]{cowbook:2002}.
For the semantics of \souffle aggregates, see the official
documentation~\cite{souffle}, which
describes how an aggregate depending on an outer-scope variable produces
a result for each valid assignment of that variable.
}
For employees \texttt{a} and \texttt{b}, the matching sales values sum
to $30$ and $40$ in \cref{fig:intro:instance}. 
Employee \texttt{c}, however, has no matching sale,
so the collection of values being aggregated is empty. The native 
{\HL{semantic conventions}} of
\SQL and \souffle differ on this case:
\[
  \operatorname{sum}_{C_{\SQL}}(\varnothing)
  =
  \texttt{NULL},
  \qquad
  \operatorname{sum}_{C_{\souffle}}(\varnothing)
  =
  0.
\]
Let
$B_{1,\SQL} \coloneqq (\mathcal{S}_1,C_{\SQL})$ and
$B_{1,\souffle} \coloneqq (\mathcal{S}_1,C_{\souffle})$.
The two expressions therefore give different query results on the same
instance $I_1$:
\[
  q_{e_1,B_{1,\SQL}}(I_1)
  \neq
  q_{e_2,B_{1,\souffle}}(I_1),
\]
or in short,
  $q_1(I_1)
  \neq
  q_2(I_1)$.
Concretely, $e_1$ returns \texttt{NULL} for employee \texttt{c}, whereas $e_2$
returns $0$. Since the two functions disagree on an admissible input, 
they are different:
\[
  q_{e_1,B_{1,\SQL}}
  \neq
  q_{e_2,B_{1,\souffle}}.
\]

Thus, two expressions may address the same \HL{information need}
in different notations and have \HL{conceptual evaluation strategies}[conceptual evaluation strategy] with
the same relational core, yet still have different
\HL{query mappings}[query mapping] under their native semantic
backgrounds.
The difference in behavior of $e_1$ and $e_2$ cannot be determined from their syntax 
alone; it arises from their different native semantic conventions
$C_{\SQL}$ and $C_{\souffle}$. If both expressions are instead
interpreted under the same compatible conventions, say
$C_{\souffle}$, then their query mappings agree:
\[
  \llbracket e_1\rrbracket_{\mathcal{S}_1,C_{\souffle}}
  =
  \llbracket e_2\rrbracket_{\mathcal{S}_1,C_{\souffle}}.
\]
Both then return $0$ for \texttt{c} on $I_1$. The difference under
native interpretation therefore comes from the different native
semantic conventions, not from the relational pattern structure.

\introparagraph{The same relational structure in different notations}
We call the common relational core underlying these conceptual
evaluation strategies the
\emph{\HL{relational pattern structure}} of an expression:
an abstraction of how the expression derives its result from
\HL{table-reference roles}[table-reference role] and relational
constructs, independent of surface notation and semantic conventions.
Thus, $e_1$ and $e_2$ instantiate the same relational pattern structure
despite their different native semantic conventions.

The next example illustrates the distinction between \HL{relation names}[relation name] and
the {\HL{roles}[table-reference role]} 
played by their occurrences
(which dissociation later makes
explicit). 
It also illustrates two further points:
($i$) equal query mappings need not imply equal pattern structures, and
($ii$) the same abstract pattern structure may occur over different
schemas.

\introparagraph{A repeated relation name}
Consider a database schema $\mathcal{S}_2$ with the single relation
schema $\mathsf{R}(\mathsf{A},\mathsf{B})$, primary key
$\{\mathsf{A},\mathsf{B}\}$, and \texttt{NOT NULL} constraints on both
attributes.
\Cref{fig:intro:rr:instance} shows an admissible
database instance $I_2$ over $\mathcal{S}_2$.

Suppose we want, for every value of $\mathsf{A}$,
the sum of the corresponding values of $\mathsf{B}$.
The \SQL expression $e_{\mathit{group}}$ 
in \cref{fig:intro:rr:query:sql:eg}
states this directly
with a \sql{GROUP BY} clause. 
\souffle has no \sql{GROUP BY} clause. Thus, to express the same
grouping task in \souffle, $e_3$ in
\cref{fig:intro:rr:query:souffle:e3} refers to $\mathsf{R}$ twice:
the outer atom \texttt{R(x,\_)} supplies the
$\mathsf{A}$ values that determine the groups, while
\texttt{R(x,y)} supplies the matching $\mathsf{B}$ values to be summed.

Under their native conventions and over the admissible instances of
$\mathcal{S}_2$, $e_{\mathit{group}}$ and $e_3$ have the same 
\HL{query mapping} 
(which we abbreviate as $q_{\mathit{group}}$ and $q_3$)
but different 
\HL{relational pattern structures}[relational pattern structure].
Expression $e_{\mathit{group}}$ uses one
\HA{table reference}
(an occurrence of a relation name in an input or binding position~\cite{DBLP:journals/sigmod/GatterbauerD25}) 
together with a grouping operator, whereas $e_3$
uses two correlated table references: the inner aggregate depends on
the outer value $x$.
Notice that these \HL{integrity constraints} are crucial for this equality of query
mappings:
the key excludes duplicate base tuples that \SQL would otherwise count
with multiplicity, and the non-null constraints keep the inputs within
\souffle's value domain.

Expression $e_4$ in \cref{fig:intro:rr:query:sql:e4} is the \SQL
counterpart of $e_3$. Its aliases \texttt{RI} and \texttt{RO} name the
inner and outer bindings, but both range over 
the same \HL{relation value} $I_2(\mathsf{R})$.
The subquery is correlated because it refers to
\texttt{RO.A} from the outer binding. Under native \SQL bag semantics,
each outer tuple produces a result tuple, so $(\texttt{a},3)$ occurs
twice. Under \souffle's set-oriented conventions, the duplicate is
collapsed. 
Thus, $e_3$ and $e_4$ have the same 
\HL{relational pattern structure} 
but different native 
\HL{query mappings}[query mapping].

\begin{figure}[t]
\setlength{\tabcolsep}{3pt}
\renewcommand{\arraystretch}{0.92}
\centering

\begin{subfigure}[t]{0.15\linewidth}
  \centering
  \begin{tabular}[t]{|c|c|}
    \multicolumn{2}{l}{\textsf{R}} \\
    \hline
    \rowcolor{gray!15}
    \underline{\textsf{A}} & \underline{\textsf{B}} \\
    \hline
    \texttt{a} & 1 \\
    \texttt{a} & 2 \\
    \texttt{b} & 3 \\
    \hline
  \end{tabular}
  \caption{$\mathcal{S}_2$, $I_2$}
  \label{fig:intro:rr:instance}
\end{subfigure}
\hspace{1mm}
\begin{subfigure}[t]{40mm}
  \centering
  \begin{minipage}[t]{\linewidth}
\begin{lstlisting}[style=sqlcode]
select R.A, sum(R.B) as sm
from R
group by R.A
\end{lstlisting}
\vspace{-2mm}
  \end{minipage}
  \caption{$e_{\mathit{group}}$ (SQL)}
  \label{fig:intro:rr:query:sql:eg}
\end{subfigure}
\hspace{2mm}
\begin{subfigure}[t]{0.3\linewidth}
  \centering
  \begin{tabular}[t]{|c|c|}
    \multicolumn{2}{l}{\textsf{Out}} \\
    \hline
    \rowcolor{gray!15}
    \textsf{A} & \textsf{sm} \\
    \hline
    \texttt{a} & 3 \\
    \texttt{b} & 3 \\
    \hline
  \end{tabular}
  \caption{$q_{\mathit{group}}(I_2)=q_3(I_2)$}
  \label{fig:intro:rr:out:e3}
\end{subfigure}

\begin{subfigure}[t]{\linewidth}
  \centering
  \begin{minipage}[t][4mm][t]{0.96\linewidth}
    \vspace*{5pt}
    \centering
    \begin{tabular}{@{}l@{}}
      \texttt{Q(x, sum y: \{R(x,y)\}) :- R(x,\_).}
    \end{tabular}
  \end{minipage}
  \caption{$e_3$ (Souffl\'e)}
  \label{fig:intro:rr:query:souffle:e3}
  \vspace{2mm}
\end{subfigure}

\begin{subfigure}[t]{37mm}
  \centering
  \begin{minipage}[t]{\linewidth}
\begin{lstlisting}[style=sqlcode]
select RO.A,
 (select sum(RI.B) as sm
  from R RI
  where RI.A = RO.A)
from R RO
\end{lstlisting}
\vspace{-2mm}
  \end{minipage}
  \caption{$e_4$ (SQL)}
  \label{fig:intro:rr:query:sql:e4}
\end{subfigure}
\hspace{2mm}
\begin{subfigure}[t]{0.25\linewidth}
  \centering
    \begin{tabular}[t]{|c|c|}
      \multicolumn{2}{l}{\textsf{Out}} \\
      \hline
      \rowcolor{gray!15}
      \textsf{A} & \textsf{sm} \\
      \hline
      \texttt{a} & 3 \\
      \texttt{a} & 3 \\
      \texttt{b} & 3 \\
      \hline
    \end{tabular}
  \caption{$q_4(I_2)$}
  \label{fig:intro:rr:out:e4}
\end{subfigure}

\caption{A schema and instance (a), a grouped \SQL expression and its
result (b, c), a \souffle expression with the same query mapping over
$\mathcal{S}_2$ (d), and a correlated \SQL expression with the same
two-role 
pattern structure as $e_3$ but a
duplicate-preserving result (e, f).}
\label{fig:intro:rr:motivating}
\end{figure}

\introparagraph{Relational pattern structure}
Each {\HL{table reference}} plays a distinct role in how the query
conceptually derives its result.
Thus, the two occurrences of $\mathsf{R}$ in
$e_3$ and $e_4$ serve as separate inner and outer
\emph{\HA{table-reference roles}[table-reference role]},
even though ordinary query evaluation supplies both with the same
\HL{relation value} $I_2(\mathsf{R})$.
The roles and their relationships form part of the structural pattern
of a query; \HL{dissociation} will later make these roles explicit by
assigning each a distinct relation name.

We write $P_e$ for the \emph{\HA{relational pattern structure}} of
expression $e$: its abstract relational structure, independent of
surface notation and semantic conventions and, up to systematic
renaming, of concrete schema names.%
\footnote{
The relational pattern structure abstracts from concrete schema vocabulary, while
its \emph{instantiations} are grounded in relation signatures, attributes, and
compatible constructs. 
Textual representations of an abstract pattern typically
still require names or explicit placeholders, while Relational
Diagrams can identify table and attribute roles by graphical
position and connections (without names or explicit placeholder tokens).}
For the four expressions,
$P_{e_1}=P_{e_2}=P_{e_3}=P_{e_4}$.
We call this common structure
\textsc{correlated subcollection summary}:
\begin{quote}
\emph{Given an outer collection and an inner collection,
use each outer tuple to determine a matching subcollection of inner tuples.
Aggregate that subcollection, and output selected attributes of the outer
tuple together with the resulting summary.}
\end{quote}
The pattern has an outer role, an inner role, a matching condition, a
copied outer value, and an aggregation step. It abstracts from surface
notation, systematic renaming, the particular aggregate function, and
semantic conventions such as empty-aggregate behavior.

The query expression $e_{\mathit{group}}$ is deliberately not among
these four: it uses one table-reference role and an explicit grouping
operator. Although $e_{\mathit{group}}$ and $e_3$ have the same query
mapping over $\mathcal{S}_2$, their pattern structures differ. 
This equality relies on the integrity constraints of $\mathcal{S}_2$; 
once they are removed, the structural difference can also become observable
through different results on suitable inputs.
We return to this point later.

\begin{figure}[t]
\centering

\begin{subfigure}[t]{\linewidth}
  \centering
  \begin{minipage}[t][3mm][t]{0.96\linewidth}
    \vspace{0pt}
    \centering
    \begin{tabular}{@{}l@{}}
      \texttt{Q(x, sum y: \{R1(x,y)\}) :- R2(x,\_).}
    \end{tabular}
  \end{minipage}
  \caption{$e'_3$ (Souffl\'e)}
  \label{fig:intro:rr:query:souffle:e3:dissociated}
  \vspace{2mm}
\end{subfigure}

\begin{subfigure}[t]{38mm} %
  \centering
  \begin{minipage}[t]{\linewidth}
\begin{lstlisting}[style=sqlcode]
select RO.A,
 (select sum(RI.B) as sm
  from R1 RI
  where RI.A = RO.A)
from R2 RO
\end{lstlisting}
\vspace{-2mm}
  \end{minipage}
  \caption{$e'_4$ (SQL)}
  \label{fig:intro:rr:query:sql:e4:dissociated}
\end{subfigure}

\caption{\HL{Dissociation}[dissociation] assigns distinct
\HL{relation names}[relation name] to the inner and outer
\HL{table-reference roles}[table-reference role].}
\label{fig:intro:rr:dissociated}
\end{figure}

\introparagraph{Dissociation and relational pattern denotation}
In ordinary evaluation of $e_3$ or $e_4$, both roles receive the same
relation value $I_2(\mathsf{R})$. \emph{\HA{Dissociation}[dissociation]}
replaces the \HL{relation name} of each table reference with a distinct
schema-level name, allowing the role inputs to vary independently.
For $e_3$ and $e_4$, this yields the \HA{dissociated query expressions}[dissociated query expression] in
\cref{fig:intro:rr:dissociated}. 
Notice the two new relation names
$\mathsf{R1}$ and $\mathsf{R2}$, which may receive independent relation
values. By contrast, the \SQL aliases \texttt{RI} and \texttt{RO} only
name local bindings and do not create independent schema inputs.

We write a \HA{database schema} $\mathcal{S}$ as a pair
$(\mathcal{R}_{\mathcal{S}},\Gamma_{\mathcal{S}})$, where
the \HA{relation-signature map} $\mathcal{R}_{\mathcal{S}}$ maps
relation names to relation signatures, and $\Gamma_{\mathcal{S}}$ is
the set of integrity constraints. 
In this notation,
$\mathcal{S}_2 = \bigl(\{\mathsf{R}:\tau_{\mathsf{R}}\},
\Gamma_{\mathcal{S}_2}\bigr)$, with relation signature
$\tau_{\mathsf{R}} =
\{\mathsf{A}:\mathsf{string},\mathsf{B}:\mathsf{int}\}$ and
$\Gamma_{\mathcal{S}_2}$ containing the key and
\texttt{NOT NULL} constraints.

The two dissociated expressions have the same constraint-free
dissociated schema:
\[
  \mathcal{S}'_{e_3}
  =
  \mathcal{S}'_{e_4}
  =
  \bigl(
    \{\mathsf{R1}:\tau_{\mathsf{R}},
      \mathsf{R2}:\tau_{\mathsf{R}}\},
    \varnothing
  \bigr).
\]
Both fresh names  $\mathsf{R1}$ and $\mathsf{R2}$ inherit the signature of $\mathsf{R}$,
and $\varnothing$ records that none of the original integrity constraints
is retained. A dissociated instance may assign them independent
relation values 
(shown here with $V_{\mathrm{inner}}$ and $V_{\mathrm{outer}}$):
\[
  I'
  =
  \{\mathsf{R1}\mapsto V_{\mathrm{inner}},
    \;\mathsf{R2}\mapsto V_{\mathrm{outer}}\}.
\]

For $B=(\mathcal{S},C)$, let
$B'_e=(\mathcal{S}'_e,C)$. The
\emph{\HA{relational pattern denotation}} is the query mapping of the
dissociated expression:
\[
  q'_{e,B}
  \coloneqq
  \llbracket e'\rrbracket_{B'_e}.
\]

Notice that the two notions provide complementary views of relational patterns: 
$P_e$ describes the pattern's table-reference roles
and their structural relationships 
independently of $C$, while $q'_{e,B}$ gives its query
mapping over independent role inputs under a fixed $C$ and without the
original integrity constraints.

The ordinary query result on $I_2$ is recovered by assigning 
(substituting)
the
original relation value to both roles:
\[
  I'_2
  =
  \{\mathsf{R1}\mapsto I_2(\mathsf{R}),
    \;\mathsf{R2}\mapsto I_2(\mathsf{R})\}.
\]
Then, for $e\in\{e_3,e_4\}$ and any fixed compatible $C$,
\[
  q_{e,(\mathcal{S}_2,C)}(I_2)
  =
  q'_{e,(\mathcal{S}_2,C)}(I'_2).
\]
Thus, dissociation preserves ordinary query evaluation when corresponding
roles receive the same value, while also allowing their values to vary
independently.

\begin{figure}[t]
\centering
\small
\setlength{\tabcolsep}{2pt}
\begin{tabularx}{\columnwidth}{@{}l l X@{}}
\toprule
$\eta$
& \HL{information need}
& information sought (often informal)
\\
$e$
& \HL{query expression}
& syntactic representation of the query
\\
$q$
& \HL{query mapping}
& function denoted by $e$
\\
\multicolumn{2}{@{}l}{$q(I)$ query result}
& result of evaluating $q$ on instance $I$
\\
\midrule
$P_e\!$
& \HL{relational pattern structure}
& abstract relational structure of $e$
\\
$e'$
& \HL{dissociated query expression}
& $e$ with distinct relation names for each table-reference role
\\
$q'$
& \HL{relational pattern denotation}
& semantics of $P_e$: query mapping of $e'$ on independent role inputs
\\
\bottomrule
\end{tabularx}
\caption{Overview of the main vocabulary and notation.}
\vspace{3mm}
\label{tab:vocabulary:overview}
\end{figure}

\begin{figure*}[t]
  \centering
  \begingroup

  \setlength{\tabcolsep}{0pt}
  \renewcommand{\arraystretch}{1.20}

  \newcommand{\querybar}[2]{%
    \cellcolor{#1}%
    \rule{0pt}{2.8ex}%
    \hspace{2pt}#2\hspace{2pt}%
  }

  \begin{tabularx}{\textwidth}{
    @{}
    >{\raggedright\arraybackslash}p{0.22\textwidth}
    @{\hspace{6pt}}
    >{\centering\arraybackslash}X p{4pt}
    >{\centering\arraybackslash}X p{4pt}
    >{\centering\arraybackslash}X p{4pt}
    >{\centering\arraybackslash}X p{4pt}
    >{\centering\arraybackslash}X
    @{}
  }
  \toprule
    & \textbf{$e_1$}
    &&
      \textbf{$e_2$}
    &&
      \textbf{$e_4$}
    &&
      \textbf{$e_3$}
    &&
      \textbf{$e_{\mathit{group}}$}
    \\
  \midrule

  {Informal \HL{information need}} $\eta$
    & \multicolumn{3}{c}{
        \querybar{gray!10}{Employee sales total}
      }
    &&
      \multicolumn{5}{c}{
        \querybar{gray!20}{Sum of $\mathsf{B}$ for each value of $\mathsf{A}$}
      }
    \\
  \addlinespace[3pt]

  {\HL{Query mapping}[query mapping] $q_{e,B}$}
    & \querybar{gray!10}{
        $q_{e_1,(\mathcal{S}_1,C_{\SQL})}$
      }
    &&
      \querybar{gray!20}{
        $q_{e_2,(\mathcal{S}_1,C_{\souffle})}$
      }
    &&
      \querybar{gray!30}{
        $q_{e_4,(\mathcal{S}_2,C_{\SQL})}$
      }
    &&
      \multicolumn{3}{c}{
        \querybar{gray!40}{
          $q_{e_3,(\mathcal{S}_2,C_{\souffle})}
          =
          q_{e_{\mathit{group}},(\mathcal{S}_2,C_{\SQL})}$
        }
      }
    \\
  \addlinespace[3pt]

  {\HL{Relational pattern structure}[relational pattern structure]} $P_e$
    & \multicolumn{7}{c}{
        \querybar{gray!10}{Correlated subcollection summary}
      }
    &&
      \querybar{gray!20}{Grouped aggregate}
    \\
  \addlinespace[2pt]

  \bottomrule
  \end{tabularx}

  \endgroup

  \caption{Relationships among the five query expressions at three
levels. Within each row, a contiguous shaded block groups expressions
that share the same information need, query mapping, or relational
pattern structure, respectively (gray shades have no meaning across
rows). Query mappings use each expression's native semantic conventions.
The equality of the query mapping between $e_3$ and $e_{\mathit{group}}$ is relative to the
admissible instances of $\mathcal{S}_2$, including its key and
non-null constraints.}
  \label{tab:intro:query-relationships}
\end{figure*}

\introparagraph{Pattern isomorphism across schemas and languages}
Dissociation exposes the inner and outer roles of all four expressions
as independently named inputs. A
\emph{\HA{pattern-preserving schema mapping}} $\lambda$ aligns the
roles and query-relevant attributes of the $\mathsf{R}$ expressions with those
of the employee-and-sales expressions:
\[
  \mathsf{R1}(\mathsf{A},\mathsf{B})
  \leftrightarrow
  \mathsf{Sales}(\mathsf{eid},\mathsf{val}),
  \qquad
  \mathsf{R2}(\mathsf{A})
  \leftrightarrow
  \mathsf{Empl}(\mathsf{eid}),
\]
together with
$\mathsf{A}\leftrightarrow\mathsf{eid}$ and $\mathsf{sm}\leftrightarrow \mathsf{sm}$ in the
output. The unused attributes $\mathsf{Sales.date}$ and
$\mathsf{Empl.k}$ do not participate in the alignment.

Informally, two expressions are
\emph{\HA{pattern-isomorphic}} if such a bijective alignment makes
their dissociated denotations equal on every aligned dissociated input
under every shared compatible choice of semantic conventions $C$.
The alignment above therefore shows that $e_1,e_2,e_3$, and $e_4$ are
pairwise pattern-isomorphic: they instantiate the same correlated
subcollection summary over different schemas and notations.

The examples distinguish three levels at which queries may be called
``the same'': they may
($i$) address the same \emph{\HL{information need}} $\eta$,
($ii$) have the same \emph{\HL{query mapping}} $q_{e,B}$ under a
semantic background $B$, or
($iii$) have pattern-isomorphic
\emph{\HL{relational pattern structures}[relational pattern structure]}.
Expressions $e_1$ and $e_2$ share the first and third properties but
have different query mappings under their native conventions.
Expressions $e_{\mathit{group}}$ and $e_3$ have the same query mapping
over $\mathcal{S}_2$ but different pattern structures.
Finally, $e_1,\ldots,e_4$ are pairwise pattern-isomorphic.

Keeping these concepts distinct is essential:
without the notion of relational pattern, debates on language design tend to focus on logical expressiveness~\cite{DBLP:journals/sigmod/GatterbauerD25,10.1145/3639316}.
And without clearly separating semantic conventions from syntax, 
comparisons may attribute to notational choices effects 
that are actually due to background context not directly encoded in a query expression.

In our framework, 
a \HL{query expression} can be seen as a derived notion:
it encodes (or instantiates) a \HL{relational pattern structure} in a \HL{notation};
under a fixed \HL{semantic background}, it denotes a \HL{query mapping}.

\section{Learning outcomes, audience, scope}

Attendees will leave with a sharper vocabulary for discussing relational query languages 
and more informed mental models of the design trade-offs between different notations. 
They will learn to recognize recurring relational patterns across different notations.
Importantly, these patterns exist independently of any particular notation.
Rather than reproducing examples chosen by recent language proposals, 
the tutorial starts from a common set of representative queries and examines how different languages express this shared workload.
Along the way, it introduces well-established concepts in relational language design (such as domain vs.\ tuple perspective, and named vs.\ positional access to relation components) and more recent terminology from the author's work (such as `inside-out' vs.\ `outside-in'). 
These concepts are illustrated first through running examples and only then formalized. 
Thus instead of starting from abstract concepts, we develop them by looking at many examples.
By the end of the tutorial, attendees should be able to look at an unfamiliar relational language, separate relational pattern from semantic conventions and notation, and explain the main design trade-offs that follow.

\introparagraph{Audience and prerequisites} 
This 3-hour tutorial targets researchers and practitioners seeking an intuitive 
guide to relational language design.
It focuses on the commonalities and differences across various languages. 
Familiarity with \SQL\ is expected. 
Basic knowledge of relational algebra (\RA) and relational calculus (\RC) is helpful but not required.

\introparagraph{Distribution} 
Slides will be made available after the tutorial on the tutorial webpage,
as with other recent tutorials by the presenter and collaborators
\cite{DBLP:journals/pvldb/Gatterbauer23,
ICDE:2024:diagrammatic:tutorial,
DBLP:conf/sigmod/TziavelisGR20,
DBLP:conf/icde/TziavelisGR22}.

\introparagraph{Relation to prior tutorials}
This 3-hour tutorial extends a prior 90-minute tutorial from SIGMOD 2026~\cite{gatterbauer2026tutorial}.
The extensions are described in the outline further below.

Recent tutorials~\cite{10.14778/3229863.3229879,
10.1145/3448016.3457545,
10.14778/3611540.3611577,
10.14778/3750601.3750698}
and surveys~\cite{DBLP:journals/csur/AnglesABHRV17} on graph query languages 
focus on graph data models and graph query languages in their own right.
Our use of graph and path notation is instead limited to a focused comparison
within a broader tutorial on \emph{relational} language design.
It remains to be seen 
which aspects of graph query languages are currently missing in relational languages, 
and whether the separation into these two different camps is really necessary. 
This raises a broader question: is the divide between relational and graph
languages partly artificial, and can a more systematic view of relational
language design help dissolve it?

\section{Tutorial Content and Outline}

The tutorial is organized around a small set of representative \SQL\ 
queries that are translated into multiple relational languages and shown side by side.
Rather than defining terminology up front, we introduce it through comparisons of concrete query expressions (e.g., contrasting domain and tuple relational calculus) and only then distill the recurring structural notation choices. 
This example-first organization matches the overall goal of the tutorial: 
to separate query mapping, relational pattern, structural notation, and semantic conventions, and to give attendees a vocabulary for recognizing the same relational structure across notationally different languages.

\introparagraph{Core concepts}
Using \ARC\ as a common reference representation and \diagrams\ as an auxiliary visual aid, we introduce the main concepts that recur across relational languages.
These include 
binding structures (tuple vs.\ domain variables),
named vs.\ positional access to relation components,
pointwise (variable-based) vs.\ point-free (tacit) notation~\cite{DBLP:journals/cacm/Backus78,DBLP:journals/toplas/Boute05,DBLP:conf/sigmod/ArefGKLMMMMNPRS25},
set vs.\ bag semantics (and list semantics, where relevant),
nested vs. pipelined composition,
declarative vs.\ procedural style,%
\footnote{%
We distinguish two notions of declarativity that should not be treated as interchangeable:
(i) \emph{membership specification}: 
a candidate is an answer exactly when it satisfies a Boolean condition.
(ii) \emph{result specification}: a conceptual computation fixes 
the complete observable result as a whole, 
but not the execution plan.
}
null handling and binary vs.\ ternary logic~\cite{DBLP:conf/pods/LibkinP23},
explicit joins via equality predicates,
implicit joins via shared variables,
joins via path expressions~\cite{DBLP:conf/vldb/FrohnLU94, Zaniolo:1983},
grouping from the inside out (FIO) vs.\ from the outside in (FOI)~\cite{CIDR:2026:ARC},
negation patterns,
correlated subqueries, scalar subqueries, lateral joins,
and derived attributes.
We relate these choices to different user tasks: composing and reading queries~\cite{DBLP:journals/csur/Reisner81}, and revising existing queries.
We also borrow concepts from programming-language design, especially orthogonality~\cite{DBLP:journals/toplas/Boute05,DBLP:journals/lisp/Steele99},
and will discuss abstraction (the art of omitting unnecessary detail).

\introparagraph{Languages}
We begin with the classical core languages 
\SQL~\cite{DBLP:journals/pvldb/GuagliardoL17}, 
tuple relational calculus (\TRC)~\cite{Disjunctions:SIGMOD:2026}, 
domain relational calculus (\DRC), 
relational algebra (\RA), and Datalog~\cite{DBLP:journals/tkde/CeriGT89}/Soufflé~\cite{souffle,DBLP:conf/cc/ScholzJSW16}, to establish the main concepts.
We discuss \SQL's conceptual evaluation strategy,
the connection to nested-loop or nested-generators in set comprehension
\cite{Gray2004}, and
illustrate comprehension-based code in both
Haskell~\cite{DBLP:conf/lfp/Wadler90,DBLP:conf/haskell/JonesW07}
and Python~\cite{python}.
We examine functional-style languages~\cite{DBLP:journals/cj/PatonG90, Gray2004,Gray2009:FQL} 
(such as DAPLEX~\cite{Shipman_1981})
and Boute's formalism~\cite{DBLP:journals/toplas/Boute05}.
We discuss Rel~\cite{DBLP:conf/sigmod/ArefGKLMMMMNPRS25} 
and Morel~\cite{morel}
as two recent relational programming languages that revisit the boundary between query language and host languages, and extend relational querying toward programming in the large.
We take a deeper look at formalisms for aggregation, starting from Klug's formalism for aggregate queries under set semantics~\cite{DBLP:journals/jacm/Klug82}, 
which influenced subsequent comprehension-based models for database programming languages (DBPLs)~\cite{DBLP:conf/dbpl/Trinder91,DBLP:journals/jiis/GrustS99,DBLP:journals/sigmod/BunemanLSTW94, DBLP:journals/tods/FegarasM00},
including extensions of logic with aggregate operators~\cite{DBLP:journals/jacm/HellaLNW01}.
Finally, we discuss recent compositional proposals such as Google's pipe syntax~\cite{DBLP:journals/pvldb/ShuteBBBDKLMMSWWY24}, SaneQL~\cite{DBLP:conf/cidr/0001L24}, and PRQL~\cite{prql}, 
asking how pipelining affects modularity and ease of reading queries.
Depending on time, we briefly connect these ideas to dataframe-style systems such as pandas~\cite{mckinney-proc-scipy-2010} 
and Ibis~\cite{ibis}
and to point-free array languages such as APL~\cite{Iverson:1980} and J~\cite{J:language}.

\introparagraph{Extensions over SIGMOD'26 tutorial}
We introduce \emph{recursion} through \Datalog\ and recursive \SQL,
covering alternatives to conventional fixpoint evaluation~\cite{DBLP:conf/cidr/HirnG23}
and recursion with aggregation~\cite{DBLP:journals/sigmod/KhamisNPSW22}.
We compare path-query notation across relational, graph, and deductive 
languages~\cite{DBLP:conf/cidr/ShaikhhaXTSH26},
including GQL and SQL/PGQ graph patterns~\cite{DBLP:conf/sigmod/DeutschFGHLLLMM22}
and GEM's \emph{functional joins}~\cite{Zaniolo:1983}.

We look beyond flat relations to nested relational data. 
Building on general models and languages for complex values~\cite{DBLP:journals/vldb/AbiteboulB95}, 
we compare SQL++~\cite{DBLP:conf/icde/0001CGOPSVW24} with nested-value and array support in \SQL. 
The examples expose differences in nesting and unnesting, 
grouping, and order (unordered collections versus ordered arrays or lists).

Finally, we survey three ways to declaratively
express NP search problems and problems at higher levels of the
polynomial hierarchy:
(\emph{i}) normal logic programs with recursion and default negation,
interpreted under stable-model semantics, a core fragment of answer-set programming
(ASP)~\cite{DBLP:journals/cacm/BrewkaET11};
(\emph{ii}) existentially quantified relation variables in
existential second-order logic (ESO), which captures NP over finite
structures, together with database-language
realizations~\cite{DBLP:journals/tplp/CadoliM07}; and
(\emph{iii}) disjunctive rule heads,
yielding disjunctive logic
programs~\cite{DBLP:journals/csur/DantsinEGV01}.
The first two support NP-level
search, whereas the third reaches the second level
of the polynomial hierarchy.

Stable-model semantics has been used in databases, for example, 
for consistent query answering over inconsistent
databases~\cite{DBLP:journals/tplp/ArenasBC03}
and trust-based conflict resolution~\cite{DBLP:conf/sigmod/GatterbauerS10}.
Related database languages and systems expose combinatorial choice and
optimization through different mechanisms, including free
second-order relation variables in
LogicBlox~\cite{DBLP:conf/amw/ArefKPV15},
solver-integrated SQL in
SolveDB~\cite{DBLP:conf/ssdbm/SiksnysP16},
package queries~\cite{DBLP:journals/vldb/BrucatoAM18},
the DeciDB prototype~\cite{deciDB}, and 
polymorphic SQL~\cite{pratten:2025}.
If time permits, we discuss how the ASP system
clingo~\cite{DBLP:journals/tplp/GebserKKS19,clingo}
permits disjunctive ASP that can automate hardness reductions~\cite{DBLP:journals/pacmmod/MakhijaG23}.

\section{Author information}

Wolfgang Gatterbauer is an Associate Professor at the Khoury College of Computer Sciences at Northeastern University. His research interests lie at the intersection of theory and practice of data management. 
He has given tutorials on visual query representations at VLDB'23~\cite{DBLP:journals/pvldb/Gatterbauer23} and ICDE'24~\cite{ICDE:2024:diagrammatic:tutorial}
and on query optimization 
at SIGMOD'20~\cite{DBLP:conf/sigmod/TziavelisGR20}
and ICDE'22~\cite{DBLP:conf/icde/TziavelisGR22}.
Slides from those tutorials are available on the tutorial pages,
including recorded videos for one of them.
The content of this tutorial is heavily informed by 
his work on visual representations of relational queries~\cite{10.1145/3639316,Gatterbauer2022PrinciplesQueryVisualization,
DBLP:journals/sigmod/GatterbauerD25,
Disjunctions:SIGMOD:2026},
by the recent proposal of an abstract relational query language that can embed other relational languages 
and represent their patterns~\cite{CIDR:2026:ARC},
and by a prior shorter tutorial on the same topic at SIGMOD 2026~\cite{gatterbauer2026tutorial}.

\begin{acks}
Thanks to 
Molham Aref,
Diandre Sabale, 
and Val Tannen
for helpful comments.
A large language model (LLM) was used to refine the wording and improve the clarity of this tutorial proposal. 
\end{acks}

\balance
\bibliographystyle{ACM-Reference-Format}

\bibliography{design-vldb-tutorial.bib}

\end{document}